\RequirePackage{ifpdf}
\ifpdf 
\documentclass[pdftex]{sigma}
\else
\documentclass{sigma}
\fi

\begin{document}

\allowdisplaybreaks

\renewcommand{\thefootnote}{$\star$}

\renewcommand{\PaperNumber}{052}

\FirstPageHeading

\ShortArticleName{Determinantal Representation of the Time-Dependent
Stationary Correlation Function}

\ArticleName{Determinantal Representation of the Time-Dependent\\
Stationary Correlation Function for the Totally\\ Asymmetric Simple
Exclusion Model\footnote{This paper is a contribution to the Proceedings of the XVIIth International Colloquium on Integrable Systems and Quantum Symmetries (June 19--22, 2008, Prague, Czech Republic). The full collection
is available at
\href{http://www.emis.de/journals/SIGMA/ISQS2008.html}{http://www.emis.de/journals/SIGMA/ISQS2008.html}}}

\Author{Nikolay M. BOGOLIUBOV}
\AuthorNameForHeading{N.M. Bogoliubov}

\Address{St.~Petersburg
Department of V.A. Steklov Mathematical Institute,\\  Fontanka 27,
St.~Petersburg 191023, Russia}
\Email{\href{mailto:bogoliub@pdmi.ras.ru}{bogoliub@pdmi.ras.ru}}

\ArticleDates{Received October 30, 2008, in f\/inal form April 14,
2009; Published online April 23, 2009}

\Abstract{The basic model of the non-equilibrium low dimensional
physics the so-called totally asymmetric exclusion process is
related to the `crystalline limit' ($q\rightarrow\infty$) of the
$SU_q(2)$ quantum algebra. Using the quantum inverse scattering
method we obtain the exact expression for the time-dependent
stationary correlation function of the totally asymmetric simple
exclusion process on a one dimensional lattice with the periodic
boundary conditions.}


\Classification{82C23; 81R50}

\section{Introduction}

The study of quantum exactly solvable models within the framework
of the quantum inverse scattering method (QISM) \cite{fad,kul,kbi}
has led to an algebraic structures which are $q$-deformations of
the universal enveloping algebras \cite{dr,frt}. These deformed
structures are  usually called now `quantum groups' or `quantum
algebras'. Integrable models   associated with the `crystalline
limit' ($q\rightarrow\infty$) \cite{jap} of these algebras play an
important role in enumerative combinatorics  \cite{sten,bres} and
non-equilibrium low dimensional physics \cite{ev,sc}.

The algebraic approach to the calculation of the correlation
functions for the integrable models within the frames of the
(QISM) was developed in the late eighties of the last century~\cite{kor}
advantage of this approach is that the form of the f\/inal answers
depends on the $R$-matrix of the model and not on the particular
realization of the operators intertwained by this matrix. It is
also important that the operator averages are calculated over the
eigenstates of the generating function of the integrals of motion.

The correlation functions for the boson model associated with the
`crystalline limit' of the $XXZ$ $R$-matrix were calculated in
\cite{bikit,biki} where it was shown that they have the
determinantal form.

The totally asymmetric simple exclusion process  (TASEP) is an
integrable model that is connected with the `crystalline limit' of
the $XXZ$ $R$-matrix. It is one of the most studied models of the
low dimensional non-equilibrium physics. The Bethe ansatz approach
to the solution of the model was developed in the papers
\cite{gwa,der,gol,gol2,ess,mal}. The relation of the model to the
QISM has been studied in \cite{noh,kim,lee}, the model with the
twisted boundary conditions within the frames of this method was
considered in \cite{nas}, and  the asymmetric exclusion process in
\cite{gol3}. The dynamical properties of TASEP are intensively
studied (see, e.g., \cite{sch,bor,sas} and the references in these
papers). The KPZ scaling limit of TASEP was considered, in
particular, in~\cite{gwa,prsp,prah}.

In the present paper we shall consider the TASEP on a periodic
lattice with the f\/inite number of sites and with the f\/inite number
of particles. We shall represent the form-factors of the local
operators in the determinantal form using the algebraic approach
of the QISM.  This  result may be considered as the main result of
the present paper.

The paper is structured as follows. In Section~\ref{section2},  the def\/inition
of the model is given. The algebraic approach to the solution of
the  model is presented in Section~\ref{section3}, in this section the
stationary states are expressed through the state vectors of the
model and the scalar product of the state vectors is expressed in
the determinantal form. In Section~\ref{section4}, the calculation of the
averages of the local projection operators over the state  vectors
of the model def\/ined on a lattice with $M$ sites is reduced to the
calculation of the scalar product of the state vectors of the
model def\/ined on the $M-1$ sites. In Section~\ref{section5}, the well known
results on the Bethe ansatz solution are presented. The answer,
based on the results of the section~\ref{section6}, for the time-dependent
stationary correlation function is obtained. Finally, conclusive
remarks are given in Section~\ref{section7}.

\section{Totally asymmetric simple exclusion model}\label{section2}

The totally asymmetric exclusion process describes a system of $N$
particles on a periodic ring with $M$ sites labelled
$i=M,M-1,\dots,2,1$. The particles move randomly only in one
direction. We shall choose this direction from right to left. The
exclusion rule forbids to have more than one particle per site.

This process can be conveniently represented using the spin
description - in each lattice site the spin-up state corresponds
to the empty site $|0\rangle $, the spin-down state corresponds to
the occupied one $|1\rangle $. The conf\/iguration of $N$ particles
($2N\leq M$) located on the sites $M\geq m_1>m_2> \dots >m_N\geq 1$ is
given by the basis vector
\[
|m_1,m_2,\dots,m_N\rangle =\sigma _{m_1}^{-}\sigma
_{m_2}^{-}\cdots \sigma _{m_N}^{-}|\Omega \rangle ,
\]
where the generating state $|\Omega \rangle $ is the state with
all spins up
\[
|\Omega \rangle =\otimes _{i=1}^M|0\rangle _i.
\]

The described process is def\/ined by a non-Hermitian Hamiltonian
\cite{gwa}:
\begin{gather}
H=-\sum_{j=1}^M\left\{ \sigma _{j+1}^{-}\sigma _j^{+}+\frac
14(\sigma _{j+1}^z\sigma _j^z-1)\right\} .  \label{hamfv}
\end{gather}
Here $\sigma ^{\pm ,z}$ are the Pauli matrices, the matrix with
subindex $j$ acts nontrivially only in the $j$-th spin space of
the total space of states of the chain $\left( {\mathbb
C}^2\right) ^{\otimes M}$: $s_j=I\otimes \cdots \otimes I\otimes
s\otimes I\otimes \cdots \otimes I$, and the periodic boundary
conditions are assumed: $\sigma _n=\sigma _{n+M}$.

In the matrix representation $-H$ is the Markov matrix that
controls the time evolution of the probability measure.

The f\/irst term of the Hamiltonian describes the hoppings of
particles
\[
\sigma _{j+1}^{-}\sigma _j^{+}|0\rangle _{j+1}|1\rangle
_j=|1\rangle _{j+1}|0\rangle _j.
\]
The second term counts the number of the allowed jumps of the
particles with hard-core repulsion.

In our paper we shall consider the time-dependent stationary correlation function~\cite{gwa}
\begin{gather}
Z_N^{-1}\langle S_N|s_1e^{-|t|H}s_m|S_N\rangle ,  \label{corff}
\end{gather}
where the projection
operator
\begin{gather}
s_k=\frac 12(1+\sigma _k^z) \label{prop}
\end{gather}
has value $1$ if there is no particle at site $k$ and $0$
otherwise. In the steady state $|S_N\rangle $ every spin
conf\/iguration with $N$ spins down has the equal weight:
\begin{gather*}
|S_N\rangle =\sum_{M\geq m_1>m_2>\dots >m_N\geq 1}|m_1,\ldots
,m_N\rangle . 
\end{gather*}
This state  is a ground state of Hamiltonian with the eigenvalue
zero:
\begin{gather*}
H|S_N\rangle =0.  
\end{gather*}
The left eigenvector of $H$ with the eigenvalue zero is
\begin{gather*}
\langle S_N|=\sum_{M\geq m_1>m_2>\dots >m_N\geq 1}\langle m_1,\ldots
,m_N|. 
\end{gather*}
The total number of conf\/igurations in the ground state
\begin{gather*}
Z_N=\langle S_N|S_N\rangle =\frac{M!}{N!(M-N)!}.  
\end{gather*}

One can express the right steady state in the following form
\cite{gsch}:
\begin{gather}
|S_N\rangle ={\cal P}_R^N|\Omega \rangle ,  \label{ststrep}
\end{gather}
where ${\cal P}_R^N$ is the $N$-th power of operator
\begin{gather*}
{\cal P}_R=\sum_{k=1}^Ms_M\cdots s_{k+1}\sigma _k^{-},  
\end{gather*}
$s_j$ are projection operators (\ref{prop}) and $\sigma_{k}^{-}$
are Pauli matrices.
 Similarly for the left steady state one has
\begin{gather}
\langle S_N|=\langle \Omega |{\cal P}_L^N  \label{ststl}
\end{gather}
with the operator ${\cal P}_L$ equal to
\begin{gather*}
{\cal P}_L=\sum_{k=1}^M\sigma _k^{+}s_{k-1}\cdots s_1.  
\end{gather*}

\section{Solution of the model}\label{section3}

To the solution of the model we shall apply the quantum inverse
scattering method \cite{fad,kul,kbi}.  The $L$-operator of the
considered model \cite{kim,nas} is a $2\times 2$ matrix with the
entries acting on the space of states of an $M$-site spin-$\frac
12$ chain:
\begin{gather}\label{lop}
L(n|u)  = \left(
\begin{array}{cc}
us_n & \sigma _n^{-} \\
\sigma _n^{+} & uI-u^{-1}s_n
\end{array}
\right)
 = ss_n+(I-s)\big(uI-u^{-1}s_n\big)+\sigma ^{-}\sigma _n^{+}+\sigma
^{+}\sigma _n^{-},
\end{gather}
where the parameter $u\in {\mathbb C}$.

The $L$-operator (\ref{lop}) satisf\/ies the intertwining relation
\begin{gather*}
R(u,v)\left( L(n|u)\otimes L(n|v)\right) =\left( L(n|v)\otimes
L(n|u)\right) R(u,v),  
\end{gather*}
in which $R(u,v)$ is the $4\times 4$ matrix
\begin{gather}
R(u,v)=\left(
\begin{array}{cccc}
f(v,u) & 0 & 0 & 0 \\
0 & g(v,u) & 1 & 0 \\
0 & 0 & g(v,u) & 0 \\
0 & 0 & 0 & f(v,u)
\end{array}
\right)   \label{r}
\end{gather}
with
\begin{gather*}
f(v,u)=\frac{u^2}{u^2-v^2},\qquad g(v,u)=\frac{uv}{u^2-v^2}.
\end{gather*}
In the quantum group theory this $R$-matrix is known as the
`crystal base' $R$-matrix \cite{jap}.

The monodromy matrix is the product of $L$-operators
\begin{gather}
T(u)=L(M|u)L(M-1|u)\cdots L(1|u)=\left(
\begin{array}{cc}
A(u) & B(u) \\
C(u) & D(u)
\end{array}
\right).  \label{mm}
\end{gather}
The commutation relations of the matrix elements of the monodromy
matrix are given by the same $R$-matrix (\ref{r})
\begin{gather}
R(u,v)\left( T(u)\otimes T(v)\right) =\left( T(v)\otimes
T(u)\right) R(u,v). \label{ttr}
\end{gather}
In the explicit form the most important of these  relations for us
are
\begin{gather}
C(u)B(v)  = g(u,v)\left\{ A(u)D(v)-A(v)D(u)\right\},  \nonumber\\
A(u)B(v)  = f(u,v)B(v)A(u)+g(v,u)B(u)A(v),  \nonumber \\
D(u)B(v)  = f(v,u)B(v)D(u)+g(u,v)B(u)D(v),  \nonumber \\
\lbrack B(u),B(v)]  = [C(u),C(v)]=0.  \label{cb}
\end{gather}

The transfer matrix $\tau (u)$ is the matrix trace of the
monodromy matrix
\begin{gather}
\tau (u)=u^{-M}{\rm tr}\,T(u)=u^{-M}\left( A(u)+D(u)\right) .
\label{trans}
\end{gather}
The relation (\ref{ttr}) means that $[\tau (u),\tau (v)]=0$ for
arbitrary values of parameters~$u$,~$v$.

The cyclic shift operator $\tau$ in the total spin space $\left(
{\mathbb C}^2\right) ^{\otimes M}$ is expressed through the
transfer matrix:
\begin{gather}
\tau \equiv \tau (1)=\Pi _{12}\Pi _{23}\cdots \Pi _{M-1M}.
\label{shift}
\end{gather}
Here
\[
\Pi _{mn}=s_ms_n+(I-s_m)(I-s_n)+\sigma _m^{-}\sigma _n^{+}+\sigma
_m^{+}\sigma _n^{-},
\]
is the permutation operator: $\Pi _{mn}\sigma _m=\sigma _n\Pi
_{mn}$. The shift operator shifts the site indices
\begin{gather*}
\tau ^{n-1}\sigma _1\tau ^{1-n}=\sigma _n  
\end{gather*}
and possesses the property $\tau ^M=I$.

In terms of the transfer matrix (\ref{trans})  the Hamiltonian
(\ref{hamfv}) is given by
\begin{gather}
H=-\frac 12\tau ^{-1}(1)\frac \partial {\partial u}\tau
(u)|_{u=1}. \label{tauham}
\end{gather}

The right state vector of the model is the vector generated by the
multiple action of operators $\tilde B(u)\equiv u^{-(M-1)}B(u)$ on
the generating state $|\Omega \rangle=\otimes _{i=1}^M|0\rangle _i
$
\begin{gather}
|\Psi (u_1,u_2,\dots,u_N)\rangle =\prod_{i=1}^N\tilde B(u_i)|\Omega
\rangle . \label{bbb}
\end{gather}
The generating state is annihilated by the operator $C(u)$%
\begin{gather*}
C(u)|\Omega \rangle =0,  
\end{gather*}
and it is an eigenvector of operators $A(u)$ and $D(u)$,
\begin{gather}
A(u)|\Omega \rangle =\alpha (u)|\Omega \rangle , \qquad D(u)|\Omega
\rangle =\delta (u)|\Omega \rangle   \label{adv}
\end{gather}
with the eigenvalues
\begin{gather*}
\alpha (u)=u^M,\qquad \delta (u)=(u-u^{-1})^M.  
\end{gather*}

The left state vector is equal to
\begin{gather}
\langle \Psi (u_1,u_2,\dots,u_N)|=\langle \Omega
|\prod_{i=1}^N\tilde C(u_i), \label{ccc}
\end{gather}
where $\tilde C(u_i)=$ $u^{-(M-1)}C(u)$, and $\langle \Omega
|B(u)=0$.

From def\/initions (\ref{lop}) and (\ref{mm}) one f\/inds by a direct
calculation that $u^{M-1}B(u)$ and $u^{M-1}C(u)$ are polynomials
in $u^2$ of degree $M-1$:
\begin{gather*}
u^{M-1}B(u)  = u^{2(M-1)}{\cal P}_R+\cdots +(-1)^{M-1}\sigma
_M^{-}s_{M-1} \cdots s_1, \\ 
u^{M-1}C(u)  = u^{2(M-1)}{\cal P}_L+\cdots
+(-1)^{M-1}s_M\cdots s_2\sigma _1^{+}. \nonumber
\end{gather*}
This decomposition allows us to express the steady states
(\ref{ststrep}) and (\ref{ststl}) through the state vectors~(\ref{bbb}) and (\ref{ccc}), namely
\begin{gather}\label{strep}
|S_N\rangle   = \lim_{\{u\}\rightarrow \infty }\prod_{i=1}^N\tilde
B(u_i)|\Omega \rangle ,   \qquad
\langle S_N|  = \lim_{\{u\}\rightarrow \infty }\langle \Omega
|\prod_{i=1}^N\tilde C(u_i).
\end{gather}

The scalar product of the state vectors (\ref{bbb}) and
(\ref{ccc}) may be evaluated directly by means of the commutation
relations (\ref{cb}) with the help of Laplace formula for the
determinant of the sum of two matrices. It also may be obtained as
a special limit of the general formula for the scalar products
(IX.6.26) of \cite{kbi}. It should be mentioned that for the
$R$-matrix (\ref{r}) the auxiliary quantum f\/ields (IX.6.24)
commute and the answer for the scalar product has the
determinantal form. It means that the scalar products of the state
vectors of the integrable models associated with the $R$-matrix
(\ref{r}) are expressed as  the determinants of the matrices with
the entries which depend on $f$ and $g$ elements of $R$-matrix and
on the eigenvalues $\alpha$ and $\delta$ (\ref{adv}) that
characterize the particular model.

For the arbitrary variables $u,v\in {\mathbb C}$ the scalar
product of the state vectors of the model is given by the
following expression
\begin{gather}
\langle \Psi (v_1,v_2,\dots,v_N)|\Psi (u_1,u_2,\dots,u_N)\rangle
\nonumber \\
\qquad {}= \left\{ \prod_{j=1}^N\frac 1{(v_ju_j)^{M-1}}\prod_{N\geq j>k\geq 1}
\frac{v_jv_k}{v_k^2-v_j^2}\prod_{N\geq l>n\geq 1}\frac{u_lu_n}{u_l^2-u_n^2}
\right\} \det Q. \label{spr}
\end{gather}

The entries of the $N\times N$ matrix $Q$ are
\begin{gather*}
Q_{jk}  = \left\{ v_j^M\big(u_k-u_k^{-1}\big)^M\left(
\frac{u_k}{v_j}\right) ^{N-1}-u_k^M\big(v_j-v_j^{-1}\big)^M\left(
\frac{u_k}{v_j}\right) ^{-N+1}\right\}
  \frac 1{\frac{u_k}{v_j}-\left( \frac{u_k}{v_j}\right)
^{-1}}. 
\end{gather*}

Using this representation and equalities (\ref{strep}) that
represent the steady states through the state vectors of the model
we can calculate the projection of the state vectors (\ref{bbb})
and (\ref{ccc}) on the steady states. For the left steady state we
have
\begin{gather} \langle S_N|\Psi (u_1,u_2,\dots,u_N)\rangle
=\lim_{\{v\}\rightarrow \infty }\prod_{i=1}^N\langle \Psi
(v_1,v_2,\dots,v_N)|\Psi (u_1,u_2,\dots,u_N)\rangle. \label{lim}
\end{gather}
This limit may be calculated with a help of the formula
\[
\lim_{v_j\rightarrow v} \frac{\det \left\{ \Phi (v_j,u_k)\right\}
}{\Delta (v)}=\det \left\{ \frac 1{j!}\frac{\partial ^j}{\partial
v^j}\Phi (v,u_k)\right\} ,
\]
where $\Phi (v,u)$ is an arbitrary  dif\/ferentiable for at least
$N$  times function of two variables, and $\Delta (v)$ is the
Vandermonde determinant. Taking the limit in (\ref{lim}) we obtain~\cite{fv}:
\begin{gather*}
\langle S_N|\Psi (u_1,u_2,\dots ,u_N)\rangle
=\prod_{k=1}^Nu_k^2\prod_{N\geq l>n\geq 1}\frac 1{u_l^2-u_n^2}\det
V^{(M)},  
\end{gather*}
where $V^{(M)}$ is a $N\times N$ matrix with the entries equal to
\begin{gather}
V_{jk}^{(M)}  = \sum_{n=0}^{j-1}(-1)^n\left(  M \atop n\right)
u_k^{2(j-1-n)},\qquad 1\leq j\leq N-1, \nonumber \\
V_{Nk}^{(M)}  = -\sum_{n=N-1}^M(-1)^n\left(  M \atop n\right)
u_k^{-2(n-N+1)}.  \label{stbeme}
\end{gather}
The projection $\langle \Psi (u_1,u_2,\dots ,u_N)|S_N\rangle $ is
given by the similar expression
\begin{gather*}
\langle \Psi (u_1,u_2,\dots ,u_N)|S_N\rangle
 = \prod_{k=1}^Nu_k^2\prod_{N\geq
n>l\geq 1}\frac 1{u_l^2-u_n^2}\det \tilde V^{(M)},  \\ 
\tilde V_{jk}^{(M)}  = \sum_{n=0}^{N-j}(-1)^n\left(  M \atop
n\right)
u_k^{2(N-j-n)},\qquad 2\leq j\leq N,  \nonumber \\
\tilde V_{1k}^{(M)} = -\sum_{n=N}^M(-1)^n\left(  M \atop n\right)
u_k^{-2(n-N+1)}.  \nonumber
\end{gather*}

\section{Form-factors}\label{section4}

The calculation of the correlation function (\ref{corff}) we start
with the calculation of the matrix elements $\langle S_N|s_1|\Psi
(u_1,u_2,\dots ,u_N)\rangle$ and $\langle \Psi
(v_1,v_2,\dots ,v_N)|s_1|S_N\rangle$ which by means of (\ref{strep})
are expressed through the form-factor of the projection operator,
namely
\begin{gather}
\langle S_N|s_1|\Psi (u_1,u_2,\dots ,u_N)\rangle
 = \lim_{\{v\}\rightarrow \infty } \langle \Psi
(v_1,v_2,\dots ,v_N)|s_1|\Psi (u_1,u_2,\dots ,u_N)\rangle ,   \nonumber\\
\langle \Psi (v_1,v_2,\dots ,v_N)|s_1|S_N\rangle
 = \lim_{\{u\}\rightarrow \infty } \langle \Psi
(v_1,v_2,\dots ,v_N)|s_1|\Psi (u_1,u_2,\dots ,u_N)\rangle .\label{ffac}
\end{gather}

By the def\/inition (\ref{mm}) the monodromy matrix may be expressed
as
\begin{gather*}
T(u)=\left(
\begin{array}{cc}
A_{M-1}(u) & B_{M-1}(u) \\
C_{M-1}(u) & D_{M-1}(u)
\end{array}
\right) \left(
\begin{array}{cc}
us_1 & \sigma _1^{-} \\
\sigma _1^{+} & uI-u^{-1}s_1
\end{array}
\right) .  
\end{gather*}
From this formula it follows that the entries of the monodromy
matrix may be expressed as
\begin{gather*}
A(u)  = uA_{M-1}(u)s_1+B_{M-1}(u)\sigma _1^{+}, \\
B(u)  = A_{M-1}(u)\sigma _1^{-}+uB_{M-1}(u)-u^{-1}B_{M-1}(u)s_1, \\
C(u)  = uC_{M-1}(u)s_1+D_{M-1}(u)\sigma _1^{+}, \\
D(u)  = C_{M-1}(u)\sigma _1^{-}+uD_{M-1}(u)-u^{-1}D_{M-1}(u)s_1.
\end{gather*}

Making use of the equalities $s_j^2=s_j$, $
s_j\sigma_j^{-}=\sigma_j^{+}s_j=0$ we obtain, in particular, that
\begin{gather}
s_1B(u)  = (u-u^{-1})B_{M-1}(u)s_1,  \qquad
C(u)s_1  = us_1C_{M-1}(u).   \label{bcrepr}
\end{gather}

Substituting the explicit form of  the state vectors (\ref{bbb})
and (\ref{ccc}) into the form-factor (\ref{ffac}) and taking into
account the commutation relations (\ref{bcrepr}) we f\/ind that for
the arbitrary values of parame\-ters~$u$, $v$ the form-factor is
proportional to the  scalar product (\ref{spr}) of the state
vectors on a lattice with $M-1$ sites:
\begin{gather*}
 \langle \Psi (v_1,v_2,\dots ,v_N)|s_1|\Psi (u_1,u_2,\dots ,u_N)\rangle  \\
 \qquad{} = \langle \Omega |\prod_{j=1}^N\tilde
C(v_j)s_1\prod_{i=1}^N\tilde B(u_i)|\Omega \rangle =\langle \Omega
|\prod_{j=1}^N\tilde C(v_j)s_1^2\prod_{i=1}^N\tilde
B(u_i)|\Omega \rangle  \\
 \qquad{} = \prod_{k=1}^N\left(1-u_k^{-2}\right)\langle \Omega
|\prod_{j=1}^N\tilde C_{M-1}(v_j)\prod_{i=1}^N\tilde
B_{M-1}(u_i)|\Omega \rangle .
\end{gather*}
Taking the limit $\{v\}\rightarrow \infty $ we obtain
\begin{gather}
 \langle S_L|s_1|\Psi (u_1,u_2,\dots ,u_N)\rangle
 = \lim_{\{v\}\rightarrow \infty }\prod_{i=1}^N\langle \Psi
(v_1,v_2,\dots ,v_N)|s_1|\Psi (u_1,u_2,\dots ,u_N)\rangle\ \nonumber \\
\qquad {} = \prod_{k=1}^N\left(1-u_k^{-2}\right)\lim_{\{v\}\rightarrow \infty
}\langle \Omega |\prod_{j=1}^N\tilde
C_{M-1}(v_j)\prod_{i=1}^N\tilde B_{M-1}(u_i)|\Omega
\rangle   \nonumber \\
\qquad{} = \prod_{k=1}^N\left(u_k^2-1\right)\prod_{N\geq l>n\geq 1}\frac
1{u_l^2-u_n^2}\det V^{(M-1)},   \label{el}
\end{gather}
where the entries of $N\times N$ matrix $V^{(M-1)}$ are (\ref{stbeme}) with $%
M$ replaced on $M-1$. Respectively we have
\begin{gather}
\langle \Psi (u_1,u_2,\dots ,u_N)|s_1|S_N\rangle
=\prod_{k=1}^Nu_k^2\prod_{N\geq n>l\geq 1}\frac 1{u_l^2-u_n^2}\det
\tilde V^{(M-1)}.  \label{ell}
\end{gather}

\section{Bethe ansatz solution}\label{section5}

By following through the arguments of the algebraic Bethe ansatz
method \cite{fad,kul,kbi} one proves that the state vectors
(\ref{bbb}) and (\ref{ccc}) are the right and the left eigenstates
of the transfer matrix~(\ref{trans}) with the same eigenvalues
\begin{gather}
\tau (v)|\Psi (u_1,u_2,\dots ,u_N)\rangle   = \Theta _N(v,\{u\})|\Psi
(u_1,u_2,\dots ,u_N)\rangle ,   \nonumber \\
\langle \Psi (u_1,u_2,\dots ,u_N)|\tau (v)  = \langle \Psi
(u_1,u_2,\dots ,u_N)|\Theta _N(v,\{u\}) \label{evtm}
\end{gather}
if allowed parameters $u_1,u_2,\dots ,u_N$ in (\ref{evtm}) satisfy
Bethe equations \cite{gwa,der,gol,gol2,ess,mal}:
\begin{gather}
\left(1-u_n^{-2}\right)^{-M}u_n^{-2N}=(-1)^{N-1}\prod_{j=1}^Nu_j^{-2}\equiv
(-1)^{N-1}U^{-2}.  \label{bethe}
\end{gather}
The eigenvalues $\Theta _N(v,\{u\})$ are equal to
\begin{gather*}
\Theta _N(v,\{u\})=\prod_{j=1}^N\frac{u_j^2}{u_j^2-v^2}+\left(1-v^{-2}\right)^M%
\prod_{j=1}^N\frac{v^2}{v^2-u_j^2}.  
\end{gather*}

From the def\/inition of the cyclic shift operator (\ref{shift}) it
follows that its eigenvalues are equal~to
\begin{gather}
\Theta _N(1,\{u\})=\prod_{j=1}^N\frac 1{1-u_j^{-2}}.
\label{evshift}
\end{gather}
From (\ref{tauham}) one f\/inds the  eigenenergies of the
Hamiltonian (\ref{hamfv})
\begin{gather*}
E_N=-\frac 12\Theta _N^{-1}(1,\{u\})\frac \partial {\partial
v}\Theta _N(v,\{u\})|_{v=1}=-\sum_{j=1}^N\frac 1{u_j^2-1}.
\end{gather*}

The obvious solution $u_1=u_2=\dots =u_N=\infty $ provides the
stationary solution with the eigenvalue $E_N=0$. The steady state
$|S_N\rangle$ is the eigenvector of the transfer matrix
(\ref{trans}) $\tau (u)$ with the eigenvalue equal to one:
\begin{gather*}
\tau (u)|S_N\rangle =\Theta _N(v,\{\infty \})|S_N\rangle
=|S_N\rangle , 
\end{gather*}
and hence it is stochastic.

The scalar product of eigenvectors (\ref{bbb}), (\ref{ccc}) is
found from formula (\ref{spr}) with $v_j=u_j$ satisfying Bethe
equations (\ref{bethe}). Understanding the diagonal elements of
the matrix $Q$ in the sense of L'H\^opitall rule one obtains the
following expression for the norm of any eigenvector:
\begin{gather}
{\cal N}^2(u_1,u_2,\dots ,u_N)  = \langle \Psi (u_1,u_2,\dots ,u_N)|\Psi
(u_1,u_2,\dots ,u_N)\rangle
 = U^{2N}\prod_{l\neq n}\frac 1{u_l^2-u_n^2}\det \tilde Q
\label{norm}
\end{gather}
with the entries of the matrix $\tilde Q$ equal to
\[
\tilde Q_{jk}=\frac{N-1+(M-N+1)u_j^{-2}}{1-u_j^{-2}}\delta
_{jk}-(1-\delta _{jk}).
\]
For the special solution $u_1=u_2=\dots =u_N=\infty $ the norm of
eigenvectors is equal to $Z_N$.

There is $Z_N$ solutions of the Bethe equations \cite{gol}, and
the state vectors belonging to the dif\/ferent sets of the solutions
of the Bethe equations are orthogonal. The eigenvectors
(\ref{evtm}) provide the resolution of the identity operator
\begin{gather*}
I=\sum_{\{u\}}\frac{|\Psi (u_1,u_2,\dots ,u_N)\rangle \langle \Psi
(u_1,u_2,\dots ,u_N)|}{{\cal N}^2(u_1,u_2,\dots ,u_N)},  
\end{gather*}
where the summation is over all dif\/ferent solutions of Bethe
equations (\ref{bethe}).

\section{Stationary correlation function}\label{section6}

For the transitionally invariant system the form-factor of the
projection operator $s_m$ is expressed through the form-factor of
the operator $s_1$:
\begin{gather*}
\langle \Psi (u_1,u_2,\dots ,u_N)|s_m|S_N\rangle   = \langle \Psi
(u_1,u_2,\dots ,u_N)|\tau ^{m-1}s_1\tau ^{1-m}|S_N\rangle  \\
\phantom{\langle \Psi (u_1,u_2,\dots ,u_N)|s_m|S_N\rangle}{}  = \prod_{j=1}^N\left( 1-u_j^{-2}\right) ^{1-m}\langle \Psi
(u_1,u_2,\dots ,u_N)|s_1|S_N\rangle ,
\end{gather*}
where the property (\ref{evshift}) was used.

The substitution of the resolution of the identity into $\langle
S_N|s_1e^{-|t|H}s_m|S_N\rangle $ gives
\begin{gather*}
 \langle S_N|s_1e^{-|t|H}s_m|S_N\rangle
 =  \sum_{\{u\}}   \frac{\langle S_N|s_1e^{-|t|H}|\Psi
(u_1,u_2,\dots ,u_N)\rangle \langle \Psi
(u_1,u_2,\dots ,u_N)|s_m|S_N\rangle }{{\cal N}^2(u_1,u_2,\dots ,u_N)}
\nonumber \\
 \phantom{\langle S_N|s_1e^{-|t|H}s_m|S_N\rangle }{} =  \sum_{\{u\}}   e^{-|t|E_N}\frac{\langle S_N|s_1|\Psi
(u_1,u_2,\dots ,u_N)\rangle \langle \Psi (u_1,u_2,\dots ,u_N)|s_1|S_N\rangle }{%
{\cal N}^2(u_1,u_2,\dots ,u_N)\prod\limits_{j=1}^N\left( 1-u_j^{-2}\right)
^{m-1}}.  
\end{gather*}

From the determinantal representations of the matrix elements
(\ref {el}), (\ref{ell}) and of the norm~(\ref{norm}) we f\/inally
obtain the answer for the stationary correlation function
(\ref{corff}):
\begin{gather*}
 \frac 1{Z_N}\langle S_N|s_1e^{-|t|H}s_m|S_N\rangle
= \left( \frac{M-N}M\right) ^2  \nonumber \\
\phantom{\frac 1{Z_N}\langle S_N|s_1e^{-|t|H}s_m|S_N\rangle =}{}  +\sum_{\{u\}}\frac{e^{-|t|E_N}}{U^{2N}}\prod_{j=1}^N\frac{%
(u_j^2-1)u_j^2}{\left( 1-u_j^{-2}\right) ^{m-1}}\frac{\det \tilde
V^{(M-1)}\det V^{(M-1)}}{\det \tilde Q}, 
\end{gather*}
where the summation is taken over all dif\/ferent solutions of Bethe
equations (\ref {bethe}) except the special one. The f\/irst term on
the r.h.s.\ is the contribution of the stationary state.

\section{Conclusive remarks}\label{section7}

The models connected with the `crystal base' $R$-matrix (\ref{r})
naturally appear  in the theory of the boxed plane partitions --
three-dimensional Young diagrams placed into a box of a f\/inite
size, and in the theory of random walkers \cite{macd,bres}. Some
aspects of these connections were discussed in the papers
\cite{b1,b2,b3,cel,uch,b4}.

In its turn plane partitions and random walkers are employed in
analyzes of the models of statistical physics describing faceted
crystals \cite{fs}, direct percolation \cite{rd}, one-dimensional
growth processes \cite{kpz}. It emphasizes the importance of the
mentioned integrable models in the theory of the non-equilibrium
processes.

\subsection*{Acknowledgements}

The work was partially supported by the RFBR grant 07-01-00358.

\newpage

\pdfbookmark[1]{References}{ref}
\LastPageEnding


\begin{thebibliography}{99}

\footnotesize\itemsep=0pt

\bibitem{fad}
Faddeev L.D., Quantum completely integrable models in field theory, {\it Soviet Sci. Rev. Sect. C: Math. Phys. Rev.}, Vol.~1, Harwood Academic, Chur, 1980, 107--155.

\bibitem{kul} Kulish  P.P.,  Sklyanin E.K., Quantum spectral transform method. Recent developments,
{\it Lecture Notes in Phys.}, Vol.~151, Springer, Berlin~-- New York, 1982, 61--119.

\bibitem{kbi}   Korepin V.E.,  Bogoliubov N.M., Izergin A.G.,
Quantum inverse scattering method and correlation functions,
{\it Cambridge Monographs on Mathematical Physics}, Cambridge University Press, Cambridge, 1993.

\bibitem{dr}  Drinfel'd V.G., Quantum groups, in Proceedings of the International Congress of Mathematicians, Vols.~1,~2 (Berkeley, Calif., 1986), Amer. Math. Soc., Providence, RI, 1987, 798--820.

\bibitem{frt}   Faddeev L.D.,  Reshetikhin N.Yu.,  Takhtajan L.A.,
Quantisation of Lie groups and Lie algebras, {\it Leningrad Math.
J.} {\bf 1} (1990), 193--225.

\bibitem{jap} Kashivara  M., Crystalizing the $q$-analogue of
universal enveloping algebras, {\it Comm. Math. Phys.} {\bf 133}
(1990), 249--260.

\bibitem{sten} Stanley  R., Enumerative combinatorics, Vol.~2, {\it Cambridge Studies in Advanced Mathematics}, Vol.~62, Cambridge University Press, Cambridge, 1999.

\bibitem{bres}   Bressoud D.M., Proofs and conf\/irmations. The story
of the alternating sign matrix conjecture, Cambridge University
Press, Cambridge, 1999.

\bibitem{ev}  Evans M.R.,  Blythe R.A.,  Nonequilibrium dynamics in low-dimensional systems, {\it Phys.~A} {\bf 313}
(2002), 110--152, \href{http://arxiv.org/abs/cond-mat/0110630}{cond-mat/0110630}.

\bibitem{sc}  Sch\"{u}tz G., Exactly solvable models for many-body systems far from
equilibrium, {\it Phase Transitions and Critical Phenomena}, Vol.~19, Editors
C.~Domb and J.L.~Lebowitz, Academic Press, San Diego, CA, 2001, 1--251.

\bibitem{kor} Korepin  V., Dual f\/ield formulation of
quantum integrable models, {\it Comm. Math. Phys.} {\bf 113}
(1987), 177--190.

\bibitem{bikit}  Bogoliubov N.M.,  Izergin A.G., Kitanine N.A.,
Correlators of the phase model, {\it Phys. Lett. A} {\bf 231}
(1997), 347--352, \href{http://arxiv.org/abs/solv-int/9612002}{solv-int/9612002}.

\bibitem{biki} Bogoliubov N.M.,  Izergin A.G., Kitanine N.A.,
Correlation functions for a strongly correlated boson system, {\it
Nuclear Phys. B} {\bf 516} (1998), 501--528, \href{http://arxiv.org/abs/solv-int/9710002}{solv-int/9710002}.

\bibitem{gwa} Gwa L.-H.,  Spohn H., Bethe solution for the dynamical-scaling
exponent of the noisy Burgers equation, {\it Phys. Rev. A} {\bf
46} (1992), 844--854.

\bibitem{der}  Derrida B.,  Lebowitz J., Exact large deviation function in the
asymmetric exclusion process, {\it Phys. Rev. Lett.} {\bf 80}
(1998), 209--213, \href{http://arxiv.org/abs/cond-mat/9809044}{cond-mat/9809044}.

\bibitem{gol}  Golinelli O.,  Mallick K.,  Bethe ansatz calculation of the
spectral gap of the asymmetric exclusion process, {\it J. Phys. A:
Math. Gen.} {\bf 37} (2004),  3321--3331, \href{http://arxiv.org/abs/cond-mat/0312371}{cond-mat/0312371}.

\bibitem{gol2} Golinelli O.,  Mallick K., Spectral gap of the totally asymmetric
exclusion process at arbitrary f\/illing, {\it J.~Phys.~A: Math.
Gen.} {\bf 38} (2005), 1419--1425, \href{http://arxiv.org/abs/cond-mat/0411505}{cond-mat/0411505}.

\bibitem{ess} de Gier J.,  Essler F.H.L., Bethe ansatz solution of the asymmetric
exclusion process with open boundaries, {\it Phys. Rev. Lett.}
{\bf 95} (2005), 240601, 4~pages, \href{http://arxiv.org/abs/cond-mat/0508707}{cond-mat/0508707}.

\bibitem{mal} Prolhac S.,  Mallick K., Current f\/luctuations
in the exclusion process and Bethe ansatz, {\it J. Phys. A: Math.
Theor.} {\bf 41} (2008), 175002, 20~pages, \href{http://arxiv.org/abs/0801.4659}{arXiv:0801.4659}.

\bibitem{noh}  Noh J.D.,  Kim D., Interacting domain
walls and the f\/ive-vertex model, {\it Phys. Rev. E} {\bf 49}
(1995), 1943--1961, \href{http://arxiv.org/abs/cond-mat/9312001}{cond-mat/9312001}.

\bibitem{kim} Kim D., Bethe Ansatz solution for crossover scaling functions
of the asymmetric $XXZ$ chain and the Kardar--Parisi--Zhang-type
growth model, {\it Phys. Rev. E} {\bf 52} (1995), 3512--3524, \href{http://arxiv.org/abs/cond-mat/9503169}{cond-mat/9503169}.

\bibitem{lee} Lee D.S.,  Kim D., Large deviation function of the partially
asymmetric exclusion process, {\it Phys. Rev. E} {\bf 59} (1999),
6476--6482, \href{http://arxiv.org/abs/cond-mat/9902001}{cond-mat/9902001}.

\bibitem{nas} Bogoliubov N.M.,  Nassar T., On the spectrum of the non-Hermitian phase-dif\/ference
model, {\it Phys. Lett. A} {\bf 234} (1997), 345--350.

\bibitem{gol3} Golinelli  O.,  Mallick K., The asymmetric simple exclusion
process: an integrable model for non-equilibrium statistical
mechanics, {\it J. Phys. A: Math. Gen.} {\bf 39} (2006),  12679--12705, \href{http://arxiv.org/abs/cond-mat/0611701}{cond-mat/0611701}.

\bibitem{sch}  Sch\"{u}tz G., Exact solution of the master equation for the
asymmetric exclusion process, {\it J. Statist. Phys.} {\bf 88}
(1997), 427--445.

\bibitem{bor} Borodin A.,  Ferrari P.,  Prahofer M.,
Sasamoto T., Fluctuation properties of the TASEP with periodic
initial conf\/iguration, {\it J. Statist. Phys.} {\bf 129} (2007),
1055--1080, \href{http://arxiv.org/abs/math-ph/0608056}{math-ph/0608056}.

\bibitem{sas}  Sasamoto T., Spatial correlations of the 1D KPZ surface on a f\/lat
substrate, {\it J. Phys. A: Math. Gen.} {\bf 38} (2005), L549--L556, \href{http://arxiv.org/abs/cond-mat/0504417}{cond-mat/0504417}.

\bibitem{prsp} Pr\"{a}hofer M., Spohn H.,
Current f\/luctuations for the totally asymmetric simple exclusion
process,  in In and Out of Equilibrium (Mambucaba, 2000), Editor V.~Sidoravicius, {\it Progr. Probab.}, Vol.~51, Birkh\"auser Boston, Boston, MA, 2002, 185--204, \href{http://arxiv.org/abs/cond-mat/0101200}{cond-mat/0101200}.


\bibitem{prah}  Pr\"{a}hofer M.,  Spohn H., Exact scaling functions for
one-dimensional stationary KPZ growth, {\it J. Statist. Phys.} {\bf
115} (2004), 255--279.

\bibitem{gsch}  Sch\"{u}tz G., Duality relations for asymmetric exclusion processes, {\it J. Statist. Phys.} {\bf 86}
(1997), 1265--1287.

\bibitem{fv}  Bogoliubov N.M., Five vertex model with the
f\/ixed boundary conditions, 
{\it Algebra i Analiz} {\bf 21} (2009), 58--78.

\bibitem{macd}   Macdonald I.G., Symmetric functions and Hall
polynomials, {\it Oxford Mathematical Monographs. Oxford Scien\-ce Publications}, The Clarendon Press, Oxford University Press, New York, 1995.

\bibitem{b1} Bogoliubov  N.M., Boxed plane partitions as an exactly
solvable boson model, {\it J. Phys. A: Math. Gen.} {\bf 38}
(2005), 9415--9430, \href{http://arxiv.org/abs/cond-mat/0503748}{cond-mat/0503748}.

\bibitem{b2}  Bogoliubov N.M., Enumeration of plane partitions and the
algebraic Bethe ansatz, {\it Theoret. and Math. Phys.} {\bf 150} (2007),
165--174.

\bibitem{b3}  Bogoliubov N.M., Four-vertex model and random tilings,
{\it Theoret. and Math. Phys.} {\bf 155} (2008), 523--535, \href{http://arxiv.org/abs/0711.0030}{arXiv:0711.0030}.

\bibitem{cel}  Tsilevich N., Quantum inverse method for the $q$-boson model, and symmetric functions,
{\it Funct. Anal. Appl.} {\bf 40} (2006), 207--217, \href{http://arxiv.org/abs/math-ph/0510073}{math-ph/0510073}.

\bibitem{uch}  Shigechi  K.,  Uchiyama M., Boxed skew plane
partition and integrable phase model, {\it J. Phys. A: Math. Gen.}
{\bf 38} (2005), 10287--10306, \href{http://arxiv.org/abs/cond-mat/0508090}{cond-mat/0508090}.

\bibitem{b4}  Bogoliubov N.M., Integrable models for vicious and friendly walkers, {\it J. Math. Sci. (N.Y.)} {\bf 143} (2007), 2729--2737.

\bibitem{fs}   Ferrari P.L.,  Spohn H., Step functions for a
faceted crystal, {\it J. Statist. Phys.} {\bf 113} (2003), 1--46.

\bibitem{rd}  Rajesh R.,  Dhar D., An exactly solvable anisotropic
directed percolation model in three dimensions, {\it Phys. Rev.
Lett.} {\bf 81} (1998), 1646--1649, \href{http://arxiv.org/abs/cond-mat/9808023}{cond-mat/9808023}.

\bibitem{kpz}  Kardar M., Parisi  G.,  Zhang Y.Z.,  Dynamic
scaling of growing interfaces, {\it Phys. Rev. Lett.} {\bf 56}
(1986), 889--892.

\end{thebibliography}
\end{document}